\documentclass[epj]{webofc}
\usepackage{hyperref}
\usepackage[varg]{txfonts}   
\usepackage[caption=false]{subfig}
\usepackage{multirow}

\begin{document}
%
%
\title{From DeepCore to PINGU} 
%
%
\subtitle{Measuring atmospheric neutrino oscillations at the South Pole}

\author{J.~P. Y\'{a}\~{n}ez\inst{1}\fnsep\thanks{\email{juan.pablo.yanez@desy.de}} for the IceCube-Gen2 Collaboration}

\institute{DESY, D-15735 Zeuthen, Germany
          }

\abstract{Very large volume neutrino telescopes (VLVNTs) observe atmospheric neutrinos over a wide energy range (GeV to TeV), after they travel distances as large as the Earth's diameter. DeepCore, the low energy extension of IceCube, has started making meaningful measurements of the neutrino oscillation parameters $\theta_{23}$ and $|\Delta m^2_{32}|$ by analyzing the atmospheric flux at energies above 10\,GeV. PINGU, a proposed project to lower DeepCore's energy threshold, aims to use the same flux to further increase the precision with which these parameters are known, and eventually determine the sign of $\Delta m^2_{32}$. The latest results from DeepCore, and the planned transition to PINGU, are discussed here.}

\maketitle
\section{Introduction}
\label{intro}
Evidence that neutrinos change flavor (oscillate) as they travel, and are thus massive, has accumulated in the last decades (see \cite{pdg2014} and references therein). Although the phenomenon has been confirmed from a variety of neutrino sources, atmospheric (anti-)neutrinos, $\nu_\mu$ and $\nu_e$, remain a very powerful tool to study it. 

The amplitude of oscillation probabilities depends on the mixing of flavor and mass eigenstates, encoded in mixing angles $\theta_{ij}$. The oscillation phase is proportional to $\Delta m_{ij}^2L/E$, where $L$ is the distance traveled by the neutrino, $E$ is its energy and $\Delta m_{ij}^2$ is the squared mass difference of the mass eigenstates. Atmospheric neutrinos are produced with energies from MeV to the TeV-scale and can travel distances as large as the Earth's diameter before detection. This means that experiments measuring these neutrinos are sensitive to the mass-splitting $\Delta m^2_{32}$ and the mixing angles $\theta_{23}$ and $\theta_{13}$ over a wide $L/E$ range \cite{vlvnt_review}.

Current VLVNTs can measure atmospheric neutrinos starting at approximately 10\,GeV \cite{antares_osc, dcdesign}, and future proposed telescopes aim to lower the threshold to a few GeV \cite{pingu, ORCA_ICRC}. Neutrinos crossing the Earth at these energies undergo neutrino-electron coherent forward scattering, which affects how they oscillate \cite{Wolfenstein:1977ue}. Between $E_\nu \approx 1-12$\,GeV, this leads to large deviations from vacuum oscillations for neutrinos or antineutrinos, depending on the ordering of the neutrino masses \cite{smirnov_resonance,param1, param2}. Above this energy range saturation occurs and, while oscillation probabilities still differ from those in vacuum, they are the same for neutrinos and antineutrinos regardless of the neutrino mass ordering.

IceCube's DeepCore \cite{dcdesign} detects neutrinos in the saturation regime. The experiment is thus suited for measuring the atmospheric parameters $\theta_{23}$ and $|\Delta m^2_\textrm{32}|$ \cite{icecube_gross, icecube_yanez}. The Precision IceCube Next Generation Upgrade (PINGU) \cite{pingu}, a proposed successor to DeepCore, will lower the energy threshold to access the resonant regime, identify the sign of $\Delta m^2_\textrm{32}$ and, with that, will place the final piece of the neutrino mass ordering puzzle.

\section{Detector performance of DeepCore and PINGU}
\label{sec:detectors}
IceCube is an ice Cherenkov neutrino detector that consists of 5160 digital optical modules (DOMs) deployed over a volume of 1\,km$^3$, at depths between 1450 and 2500\,m at the geographic South Pole \cite{icecube}. The DOMs are arranged in 86 strings, with each string holding 60 of them. Most strings are situated over a quasi-hexagonal grid, with an inter-string separation of 125\,m, and a vertical spacing between DOMs of 17\,m. Towards the middle of the detector, eight strings are located in between the regular grid, reducing the inter-string spacing to $40-70$\,m. The DOMs on these strings have 35\% higher quantum efficiency than the standard DOMs and are only 7\,m apart. In this volume, known as DeepCore, it is possible to detect neutrinos with energies as low as 10\,GeV \cite{dcdesign}.

PINGU will consist of 40 additional strings deployed inside the DeepCore volume with an inter-string spacing of approximately 20\,m \cite{pingu, hanson_vlvnt}. Each string will host 96 high quantum efficiency DOMs 3\,m apart. The additional optical modules result in a factor of 10 increase in photocathode area density with respect to DeepCore. The closer spacing between the DOMs in PINGU lowers the energy threshold for detection to a few GeV, and improves the reconstruction of the energy and arrival zenith angle of the neutrinos interacting in its volume.

\subsection{Signal and background}

There are eight distinct neutrino-nucleon interactions that DeepCore can, and PINGU will, observe: charged current (CC) $\nu_e$, $\nu_\mu$ and $\nu_\tau$, and neutral current (NC) interactions of all flavors, both of neutrinos and antineutrinos. Only two of them, CC~$\nu_\mu$ and $\bar{\nu}_\mu$, are expected to produce muons with a range similar to the DOM spacing\footnote{ While 18\% of CC~$\nu_\tau$ interactions will also produce muons from tau decay, the expected number of events of this type at energies relevant for atmospheric neutrino oscillations is negligible.}. All of these interactions below $E_\nu \leq 100$\,GeV constitute the signal for oscillation measurements in DeepCore and PINGU. Muon neutrino disappearance is the strongest effect expected, and since most muon neutrinos oscillate into tau neutrinos, $\nu_\tau$ appearance is also testable. Transitions involving electron neutrinos are relevant in the resonance region. The identification and reconstruction of these events are explained in the following sections.

Atmospheric muons, produced in the same cosmic ray showers as the neutrinos, are a dominant source of background events. In DeepCore data analyses, atmospheric muons are rejected by a variety of veto algorithms that try to correlate the signal observed in DeepCore with the arrival time of photons observed outside the DeepCore volume. The same algorithms are also used to extract a pure atmospheric muon sample from data. These data are used as a background template in the final fit of the simulation. A similar strategy is being implemented for PINGU. After the veto information has been exhausted, the remaining atmospheric muon background is rejected by requiring the events to have a minimum of 5 direct photons (see below) and by applying cuts on the reconstruction quality. 

Background events are also introduced by PMT thermal noise and radioactive decays in the glass of the DOM, which become relevant when looking for faint events. Events produced by this noise are also removed by quality cuts on the reconstruction. A more important effect of this background is that it can bias the energy reconstruction and affect the efficiency of veto algorithms. In order to model it correctly, noise simulation is tuned to the data \cite{larson}.

\subsection{Reconstruction of neutrino events}
\label{sec:reco}
All neutrino events are reconstructed using the general hypothesis that at the point of interaction a hadronic shower is produced together with a collinear muon. A muon hypothesis is fit to the event with track lengths ranging from zero (i.e. ultimately a cascade-like event) to the detector's size. Only a handful of the photons emitted in low energy neutrino interactions are observed by DeepCore, and these photons may scatter multiple times before being detected. The latest DeepCore results \cite{icecube_yanez} relied on finding unscattered photons to aid the event selection and event reconstruction. Unscattered, or direct, photons are identified by restricting the patterns formed by their arrival time to a string. While requiring 5 or more direct photons results in a 30\% signal efficiency, this criterion keeps the subset of the detected neutrinos whose direction can be easily reconstructed. The muon track direction is fit using these photons, following \cite{antares_bbfit}. Since the light produced by a cascade is also preferentially emitted in the Cherenkov angle events without a muon are also fit, although the resolution obtained is significantly worse. The neutrino interaction position, the energy of the hadronic cascade at the interaction point, and the range of the muon produced are fit in a second step using all photons recorded by the detector while the direction along which these parameters are varied is kept fixed. At $E_\nu = 15$\,GeV, the resulting median zenith angle resolution is 10$^\circ$ for $\nu_\mu$ and 25$^\circ$ for $\nu_e$, while the median energy resolution is 25\% for both.

A new method for reconstructing neutrino interactions has been successfuly developed in the last few years. It uses a maximum likelihood estimator where the expected time of arrival of photons to a DOM given a particle hypothesis is finely binned. The effects of the optical properties of the South Pole ice are taken into account. In contrast to the method described above, this approach has the advantages of being able to reconstruct nearly all neutrino events (no direct photons are required) and of producing a significantly better directional reconstruction for events without a muon present. Other resolutions obtained are comparable. The maximum likelihood reconstruction is applied to study the performance of PINGU, where the energy and zenith angle resolutions are expected to improve by approximately 30\%.


\subsection{Particle identification}
Muons with long tracks are used to identify CC~$\nu_\mu$ interactions. In the latest results from DeepCore, neutrino-induced muons are identified by comparing the quality of the muon track directional reconstruction to a cascade fit that assumes the light observed has been randomized by scattering and travels spherically outwards from its emission point. The classification stabilizes at $E_\nu \geq 30$\,GeV, and above this energy it correctly identifies 60\% of CC~$\nu_\mu$, while misidentifying 30\% of cascades. A similar strategy is being developed for the maximum likelihood reconstruction.

The particle identification in PINGU is more complex, as $\nu_e$ and $\nu_\mu$ are expected to contribute to the sensitivity to discriminate the neutrino mass ordering \cite{pingu}. A multivariate method has been put in place to classify the events according to their topology taking advantage of the reduced inter-string spacing. The classifier shows a stable behavior at $E_\nu \geq 15$\,GeV, correctly identifying about 80\% of CC~$\nu_\mu$ and misidentifying less than 20\% of cascade-like events.

\section{Analysis methods and systematic uncertainties}

Current and future analysis of DeepCore data are aimed at measuring the atmospheric neutrino flux and extracting the oscillation parameters $\theta_{23}$ and $|\Delta m^2_{32}|$. The simulation is fit to the data under the assumption of normal and inverse mass ordering, and both results are reported. The CP-violating phase $\delta$ is held to zero. Only marginal sensitivity to the neutrino mass ordering is expected from DeepCore data. A main goal of PINGU analyses, on the other hand, is to discriminate between the two possible mass orderings. In this measurement, the atmsopheric oscillation parameters would also be measured while the solar parameters and $\theta_{13}$ would be taken from global analyses of all available oscillation data \cite{Gonzalez-Garcia:2014bfa}.

A likelihood ratio method is used both for DeepCore and PINGU to find the best fit between data and simulation, and to estimate the error of the measurement. In DeepCore the method relies on the production of several high statistics simulation sets that span the space of the systematic uncertainties. Sources of these systematic uncertainties are implemented as nuisance parameters and the data is left to select the set of parameters in the simulation that describe it best. The uncertainties on the oscillation parameters are estimated from scanning the likelihood landscape around the best fit point and using a $\chi^2$ approximation. The same method is used for studies which aim at improving the precision of current DeepCore results.

\begin{table}[bt]
\small
\centering
\caption{List of the sources of error included in neutrino oscillation analyses in the latest DeepCore result (DC), ongoing studies of DeepCore data (DC+) and PINGU (P). DIS stands for deep inelastic scattering. BY refers to the six parameters used in the Bodek-Yang model implemented in GENIE.}
\label{table:uncertainties}   
\begin{tabular}{c|c|c|c|cccc}
\hline
\multicolumn{2}{c}{Source of error} & Nominal value & Uncertainty&DC & DC+ & P  \\\hline
\multirow{6}{*}{\parbox{1.7cm}{\centering Neutrino interactions}} & Total cross section scaling &\multirow{6}{*}{GENIE model \cite{genie}} & Free& x & x & x \\
& Linear energy dependence & & $E^{\pm 0.03}$& x & x & x \\
& DIS low Q$^2$ tuning & & BY $\pm[25,40]\%$ & &x&x \\
& NC scaling & & $\pm10$\% & &x&x\\
& Quasi-elastic axial mass & & $-15\%\,+25\%$ &x & x & x \\
& Resonance axial mass  & & $\pm20\%$ &x & x & x \\ \hline
\multirow{5}{*}{\parbox{1.7cm}{\centering Atmospheric $\nu$ and $\mu$ flux}} & Overall scaling &\multirow{4}{*}{Honda 2015 \cite{honda2015}} & Free &x & x & x \\
& Spectral index & & $E^{\pm 0.04}$ &x & x & x \\
& Flux ratio $\nu/\bar{\nu}$ & & $\pm$20\% & &x&x\\
& $\nu_e/\nu_\mu$ relative scaling  & & $\pm3\%$ &x & x & x \\
& Atm. $\mu$ contamination & From data  & Free & x&x&\\ \hline
\multirow{4}{*}{\parbox{1.7cm}{\centering Detection process}} & DOM overall efficiency & Muons and flashers & $\pm$10\%& x&x&\\
& DOM angular acceptance &Flashers, laser & Up to 50\% & x&x& \\
& Bulk ice model & Flashers & Models in \cite{ice_mie, ice_lea}& x&x&\\
& Hadronic energy scaling &Geant4 \cite{geant4} & $\pm$5\%& & x& x \\ \hline
\end{tabular}
\end{table}

PINGU uses two methods for estimating its sensitivity to the neutrino mass ordering. The likelihood ratio approach is implemented as outlined in \cite{franco_nmh}. In this method, pseudo-experiments are drawn from two simulation templates, each produced under a different mass ordering. The oscillation parameters requried for the two templates are selected so that they result in the case of ``maximum confusion'' between hierarchies. Each pseudo-experiment is fit under the assumption of both hierarchies and the likelihood ratio is calculated. The separation between the resulting distributions serves as an estimate of the confidence with which the two hypotheses can be discriminated. The second method used in PINGU is a simplified $\Delta \chi^2$ approach. This is a fast parametric analysis of the Fisher information matrix \cite{fisher_matrix} and relies on the partial derivatives of the event counts in each bin with respect to the parameters under study. The results obtained from the simplified $\Delta \chi^2$ and the likelihood ratio approaches have been found to be in agreement, and both are used to report the projected sensitivity of PINGU.

In all methods discussed, sources of uncertainty are implemented as additional parameters in the fit to the data. Table~\ref{table:uncertainties} contains a list of the uncertainties considered for the latest DeepCore result, those being studied for future DeepCore measurements, and those included in the PINGU studies. An additional fine-grained study of the impact of the uncertainties of the neutrino flux predicions for PINGU analyses, not listed in Table~\ref{table:uncertainties}, can be found in \cite{joakim_vlvnt}. We note that, directly impacting the measurement of the oscillation parameters, simulation tests indicate that the detector energy scaling is strongly correlated with $|\Delta m^2_{32}|$. The value of $\sin^2\theta_{23}$ shows a dependence on the assumed angular acceptance of the DOM and also on the parameters that modify the angular flux of atmospheric neutrinos. The sensitivity to the neutrino mass ordering depends strongly on the true value of $\sin^2\theta_{23}$.

\section{Results and projections for DeepCore and PINGU}

\begin{figure}[htb]
  \sidecaption
  \centering
    \includegraphics[width=0.45\columnwidth]{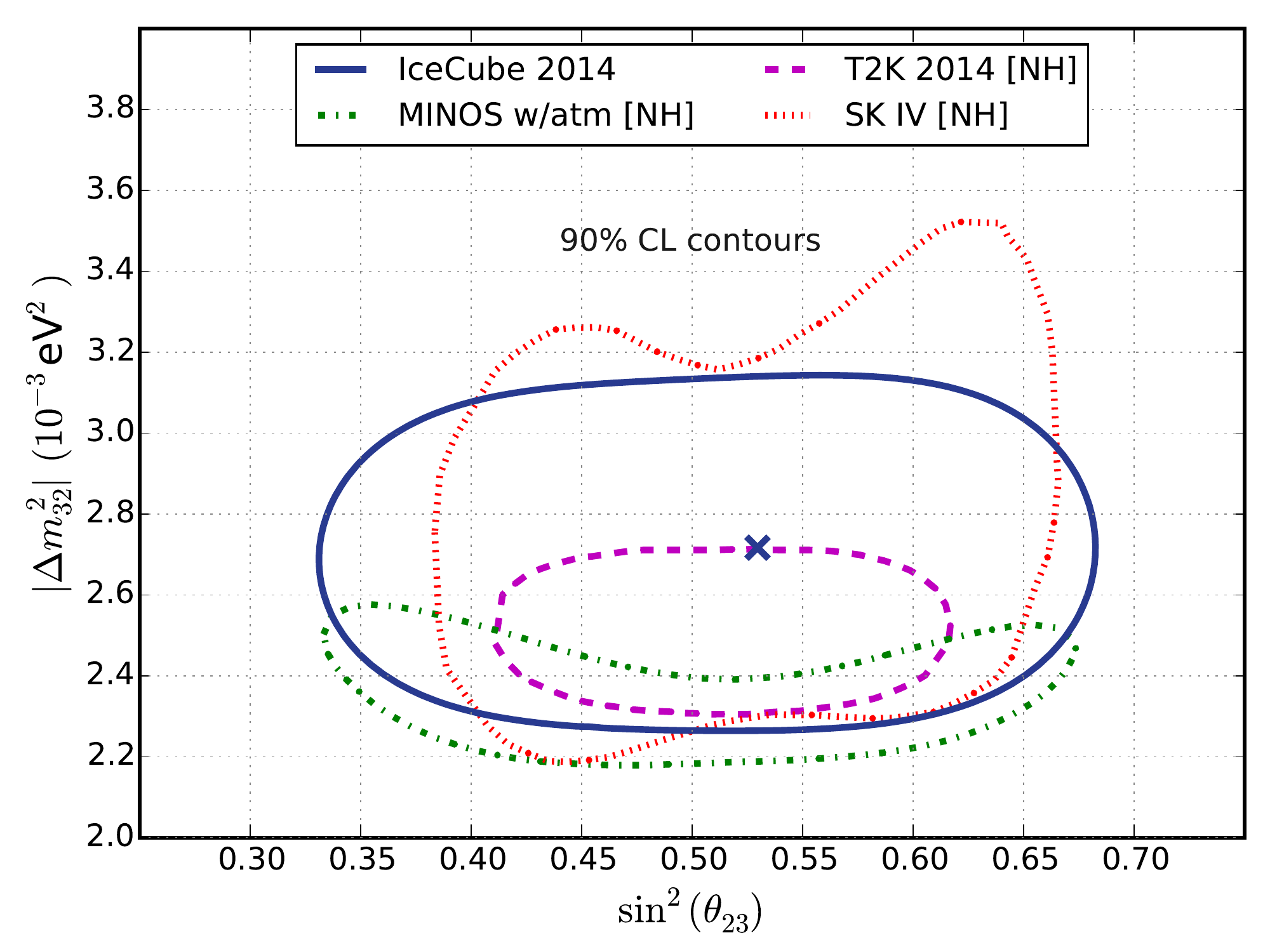}
  \caption{Comparison of best fit and confidence intervals of atmospheric oscillation parameters obtained with DeepCore \cite{icecube_yanez} with those of SuperKamoikande \cite{SK2014}, T2K \cite{T2K_latest} and MINOS \cite{minos1}.}
  \label{fig:dc_results}  
\end{figure}

The latest published results of DeepCore are a muon disappearance analysis of three years of data considering only track-like events with $\cos\theta_\mathrm{reco} < 0$ and $E_\mathrm{reco}=[6,56]$\,GeV. Some 5174 events are found in that range. The best fit point is found at $\sin^2\theta_{23} = 0.53^{+0.09}_{-0.12}$ and $\Delta m^2_{32}=2.72^{+0.19}_{-0.20}\times 10^{-3}$\,eV$^2$. Figure~\ref{fig:dc_results} shows the best fit point and the 90\% confidence regions as a function of the atmospheric oscillation parameters.

The sensitivity of an eventual re-analysis of the three years data set has been investigated in a simulation study. This study includes a less restrictive event selection (including cascade-like events), events from the complete zenith range, and the maximum likelihood event reconstruction described in Sec.~\ref{sec:reco}. The senstivity calculation using these modifications indicates the possibility to reduce the uncertainty on $|\Delta m^2_{32}|$ to half of the currently reported value, and that the uncertainty on $\sin^2\theta_{23}$ may be decreased by 20\%.

\begin{figure}[htb]
  \centering  
  \begin{subfloat}
    \centering
    \includegraphics[width=0.44\columnwidth, trim=0 0 0 0 ]{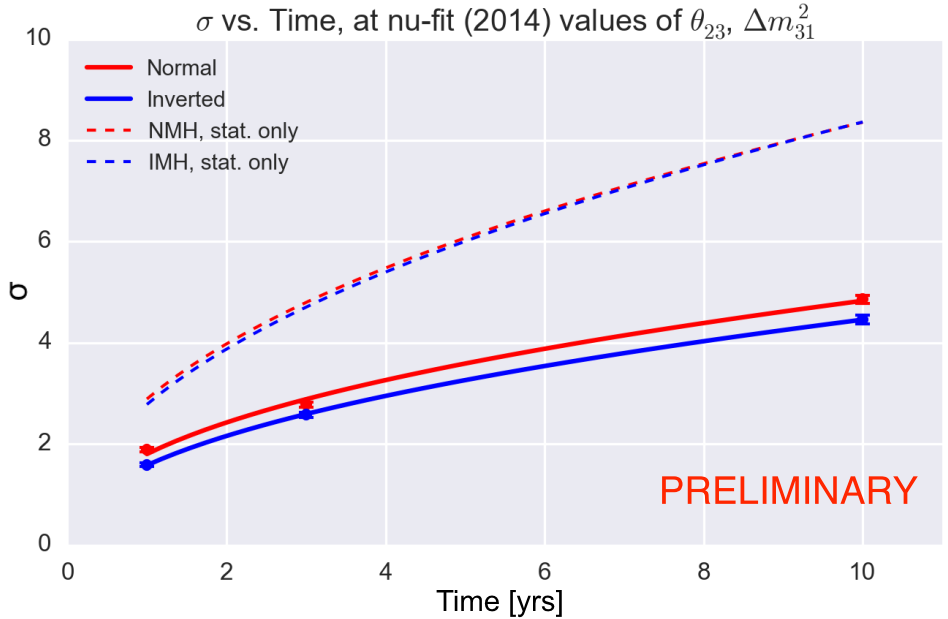}
  \end{subfloat}
  \hspace{0.3cm}
  \begin{subfloat}
    \centering
    \includegraphics[width=0.465\columnwidth, trim = 0 27 0 0]{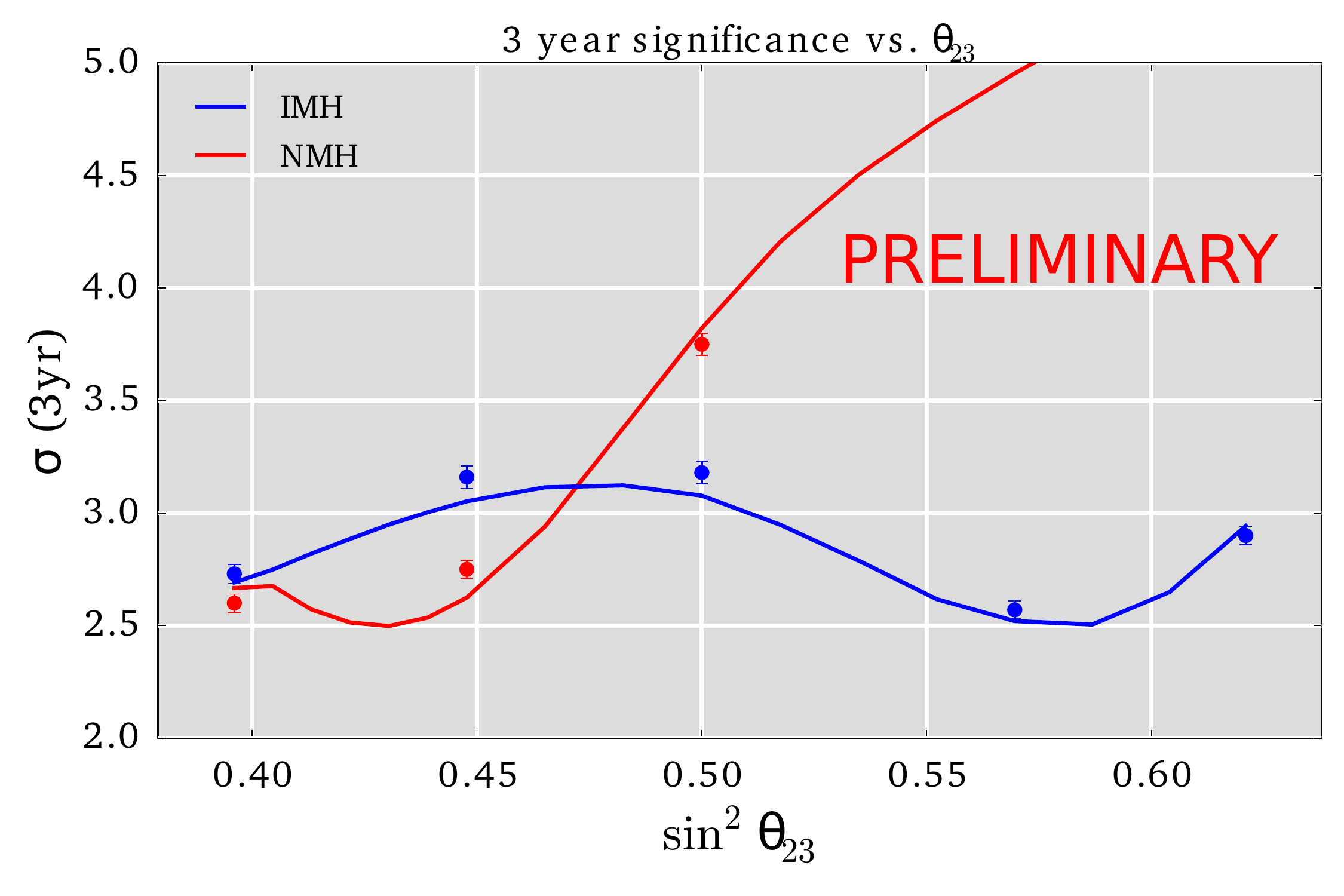}
  \end{subfloat}
  \caption{PINGU projections. Left: Expected significance for discriminating the mass ordering as a function of time. Right: Significance as a function of the true value of $\sin^2\theta_{23}$ at the three year benchmark.\label{fig:pingu}}
\end{figure}

The predicted sensitivity of PINGU to the neutrino mass ordering as a function of time is shown in Fig.~\ref{fig:pingu}~(left), where the accumulated impact of the systematic uncertainties considered is also shown. These studies indicate that PINGU will be able to identify the neutrino mass ordering with a significance of 3$\sigma$ after $3-3.5$ years of operation. The sensitivity dependence on the true value of $\sin^2\theta_{23}$ at the 3$-$years benchmark is shown in Fig.~\ref{fig:pingu}~(right). A favorable combination of true physics parameters results in a faster identification of the mass ordering.

\section{Outlook}

Atmospheric neutrinos are a valuable tool to study neutrino oscillations. IceCube's DeepCore has demonstrated that very large volume neutrino telescopes can use atmospheric neutrinos to make meaningful contributions to the field. Current results have started to approach the precision of dedicated experiments. Studies indicate that the full potential has not been reached and the addition of the PINGU array can grealty improve on these measurements by accumulating large samples of very well reconstructed GeV-scale neutrinos and ultimately identify the neutrino mass ordering.

%

%

\end{document}